\documentclass[a4paper]{jpconf}
\usepackage{graphicx}
\begin{document}
\title{An overview of (selected) recent results in finite-temperature lattice QCD}

\author{Alexei Bazavov}

\address{Physics Department, Brookhaven National Laboratory, Upton, NY 11973, USA}

\ead{obazavov@quark.phy.bnl.gov}

\begin{abstract}
I discuss recent results on lattice QCD calculations with the main emphasis on the
thermodynamics of the crossover region, restoration of the chiral
symmetry and fluctuations of conserved charges as indicator of
deconfinement, that may also be used to determine the chemical
freeze-out conditions in heavy-ion collision experiments.
\end{abstract}

\section{Introduction}

It has been established by lattice QCD calculations that
at the physical values of light quark masses and at vanishing
chemical potential there is no genuine phase
transition in QCD, but rather a ``rapid''
crossover~\cite{Bernard:2004je,Cheng:2006qk,Aoki:2006we}.
While the low-temperature
phase exhibits confinement and breaking of chiral symmetry, at high
temperatures the behavior of the theory is qualitatively different --
the interaction between quarks and gluons decreases due to asymptotic
freedom, leading to deconfinement, and the chiral symmetry is 
restored (see \cite{Petreczky:2012rq} for a recent review).
In the crossover region the QCD partition function does not exhibit
a singularity, so it is not a surprise that different physical observables
show a change in their temperature dependence at somewhat different temperatures.

\section{Restoration of chiral symmetry and $T_{pc}$ in QCD}
\begin{figure}[h]
\begin{minipage}{0.495\textwidth}
\includegraphics[width=\textwidth]{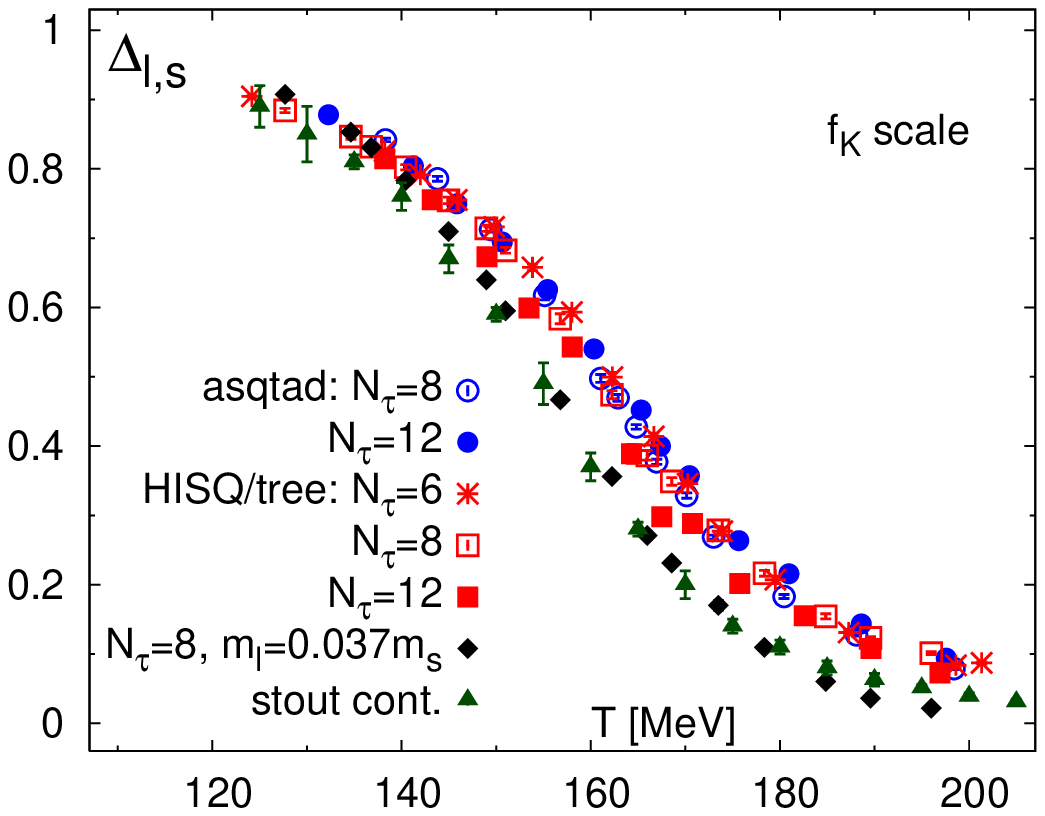}
\vspace{-9mm}
\caption{\label{fig_delta_ls}The renormalized quark condensate with the asqtad
and HISQ/tree action.}
\end{minipage}\hfill
\begin{minipage}{0.465\textwidth}
\includegraphics[width=\textwidth]{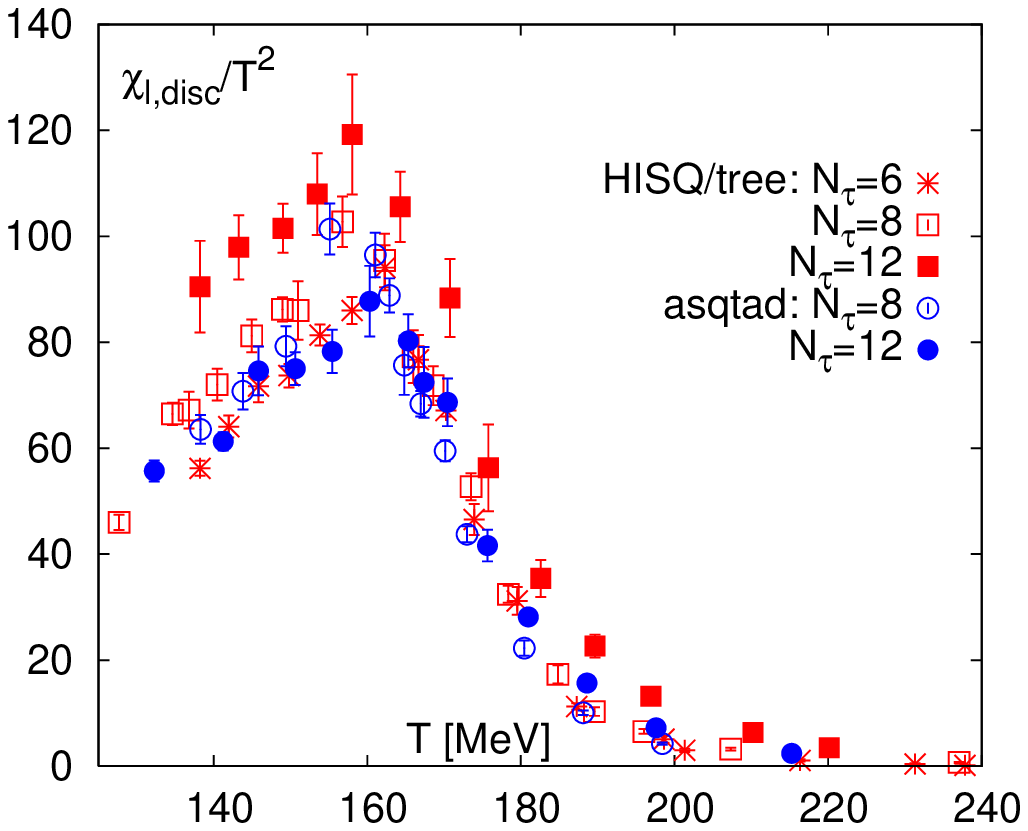}
\vspace{-9mm}
\caption{\label{fig_chi_disc}The disconnected susceptibility with the asqtad
and HISQ/tree action.}
\end{minipage} 
\end{figure}

Let us start the discussion with the phenomenon of chiral
symmetry restoration. The QCD Lagrangian at zero light quark mass allows for
independent left and right rotations, {\it i.e.} possesses a
$SU(2)_L\times SU(2)_R$ symmetry. This symmetry is however spontaneously
broken in the vacuum, as indicated by a condensate of quark anti-quark pairs.
On general grounds~\cite{Pisarski:1983ms} it is expected
that at zero light quark mass there is a second-order phase transition
in QCD (in the $O(4)$ universality class) for which the chiral condensate is
the order parameter\footnote{This is true under an additional assumption 
that the $U(1)_A$ axial symmetry breaking is still large
at the chiral critical temperature, $T_c$, and the symmetry is restored
at higher temperatures. Otherwise
the symmetry group would be enlarged, and the $O(4)$ universality class would be no longer
appropriate.}.
The renormalized quark condensate $\Delta_{ls}$ (I reserve
the name ``chiral condensate'' for the light quark condensate at zero mass) is
shown in Fig.~\ref{fig_delta_ls} and defined as:
\begin{equation}
\Delta_{ls}=\frac{\langle \bar\psi_l\psi_l\rangle_T-\frac{m_l}{m_s}
\langle \bar\psi_s\psi_s\rangle_T}
{\langle \bar\psi_l\psi_l\rangle_0-\frac{m_l}{m_s}
\langle \bar\psi_s\psi_s\rangle_0},\,\,\,
\langle \bar\psi_q\psi_q\rangle = \frac{T}{V}\frac{\partial\ln Z}{\partial m_q}
=\frac{1}{N_s^3N_\tau}\left\langle {\rm Tr} M^{-1}_q\right\rangle,\,\,\,q=l,s,
\end{equation}
where $M_q$ is the fermion matrix, \textit{i.e.} the discretized version
of the Dirac operator, $N_s^3N_\tau$ is the four-dimensional lattice volume.
The disconnected susceptibility which is proportional to fluctuations
of the quark condensate
\begin{equation}
\chi_l^{disc}=\frac{1}{N_s^3N_\tau}\left\{
\left\langle \left({\rm Tr} M^{-1}_l\right)^2 \right\rangle -
\left\langle {\rm Tr} M^{-1}_l\right\rangle^2
\right\}
\end{equation}
is shown in Fig.~\ref{fig_chi_disc}.
The results are obtained with the asqtad and HISQ/tree
action at the light to strange quark mass ratio $m_l=m_s/20$ (somewhat heavier but
close to the physical $m_l=m_s/27$).

If the physical light quark mass is small enough
one may hope that remnants of criticality (plus subleading corrections) still
govern the crossover region. Applicability of the critical scaling was
extensively studied in Ref.~\cite{Ejiri:2009ac}
with staggered fermions and the
scaling was indeed observed. (An additional complication for staggered
fermions is that at non-zero lattice spacing the relevant universality
class is $O(2)$ rather than $O(4)$. Luckily, the numerical difference between
$O(2)$ and $O(4)$ fits is not significant for our discussion, comparable with
other uncertainties, and we refer to this analysis as $O(N)$ scaling.)
This approach has been taken by the HotQCD collaboration, who 
calculated the light quark condensate and its susceptibility with 
the asqtad and HISQ/tree actions on lattices with the temporal extent
$N_\tau=6$, $8$ and $12$ at several values of the light quark mass and 
then performed fits to $O(N)$ scaling functions complemented by non-singular
terms. The pseudo-critical temperature, $T_{pc}$, defined this way
as a location of the peak of the chiral susceptibility, reduces to the
true critical temperature in the chiral limit. The result in the
continuum limit at the physical light quark mass is 
$T_{pc}=154(9)$~MeV~\cite{Bazavov:2011nk}. This is compatible with
earlier results by the Budapest-Wuppertal
collaboration that are in a range $T_{pc}=147-157$~MeV, depending
on what chiral observable is picked to determine
$T_{pc}$~\cite{Borsanyi:2010bp}.

\begin{figure}[h]
\vspace{-0.6cm}
\begin{tabular}{ccc}
\parbox{0.33\textwidth}{
  \includegraphics[width=0.32\textwidth]{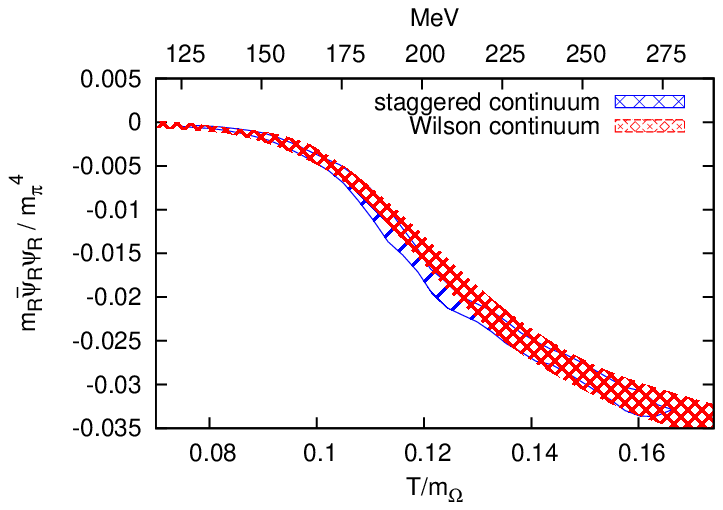}
}
&
\parbox{0.33\textwidth}{
  \includegraphics[width=0.32\textwidth]{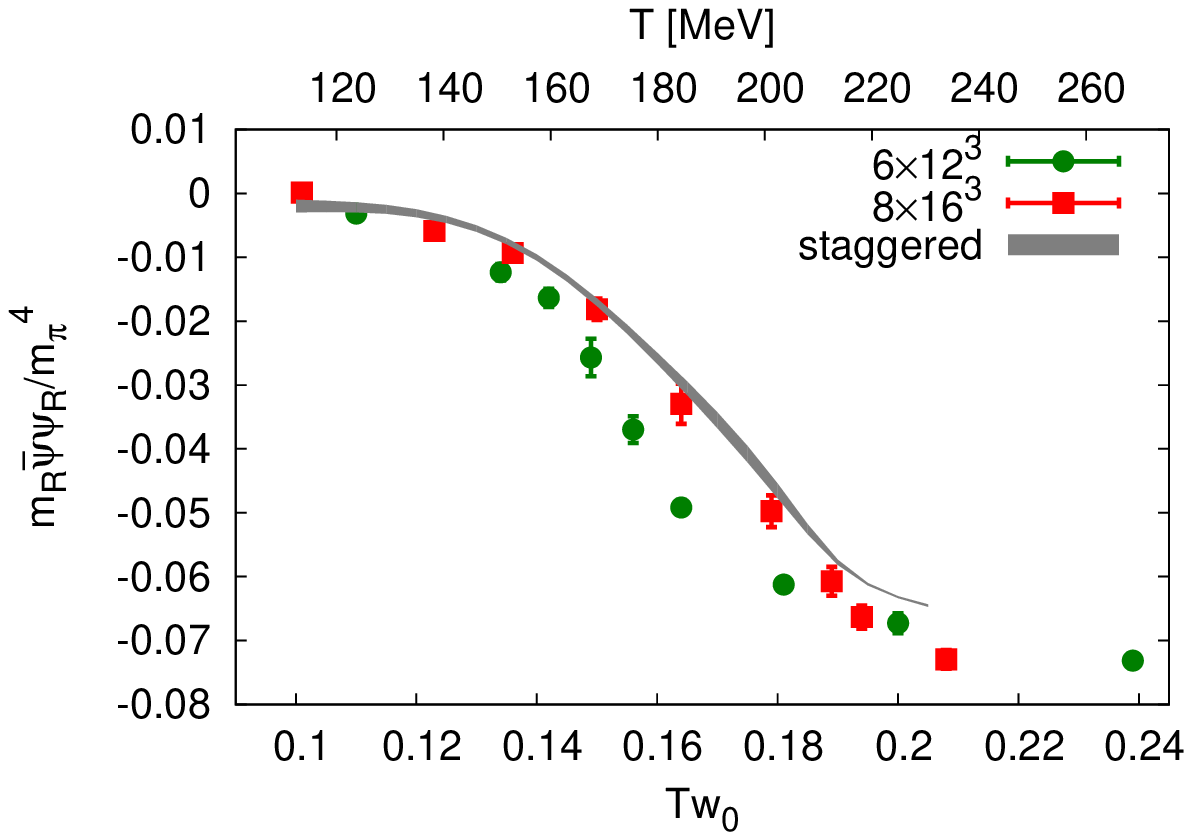}
}
&
\parbox{0.33\textwidth}{
  \includegraphics[width=0.24\textwidth,angle=-90]{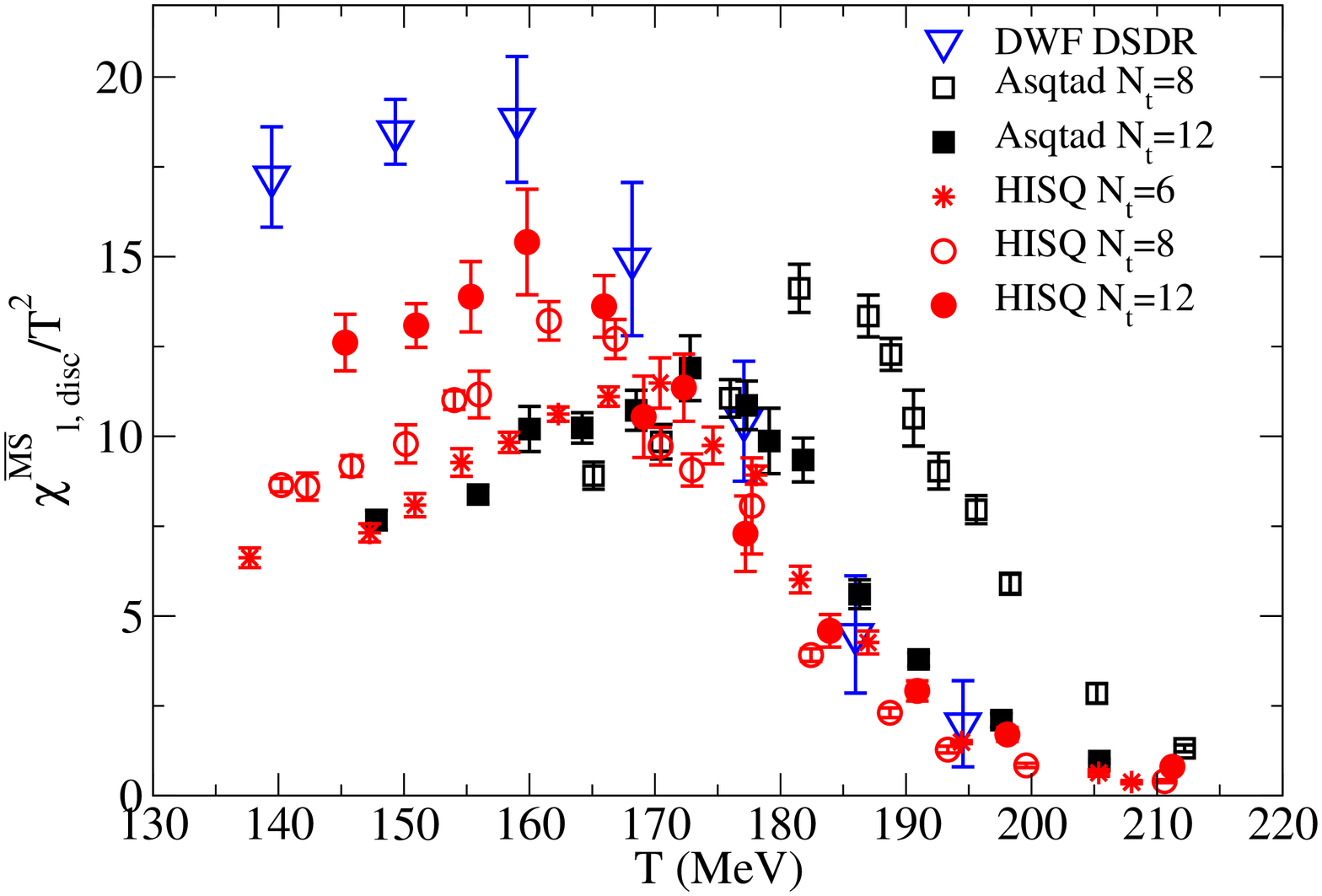}
}
\end{tabular}
\vspace{-5mm}
\caption{\label{fig_pbp_check}Comparison of the observables indicating restoration
of the chiral symmetry between staggered and three other fermion formulations (Wilson,
overlap and domain-wall). See text for details.}
\vspace{-0.6cm}
\end{figure}

The studies discussed so far rely on staggered fermions (one of the numerically
cheapest lattice fermion formulations). The fermion doubling problem leads to
16 species of lattice fermions per one in continuum and staggered fermions
reduce them to 4 species. The latter are however non-degenerate (and are labeled
by ``taste'' to distinguish from flavor). In other words, taste symmetry
is broken, for particular actions we use at $O(a^2)$, $a$ being the lattice spacing.
This leads to a distorted hadron spectrum, and taste-breaking effects have 
been identified as the largest source of systematic errors in staggered
simulations (a discussion in relation to thermodynamics is presented
in Ref.~\cite{Bazavov:2010pg}).
It is desirable to have independent tests of the results also
with other types of lattice fermions. The latter
require more computational resources. Calculations directly at the
physical mass thus may not be feasible. However, some
conclusions can be drawn from simulations at heavier masses.
The Budapest-Wuppertal collaboration showed agreement between
the continuum-extrapolated
light quark condensate for staggered and Wilson fermions at the pion
mass $m_\pi=545$~MeV~\cite{Borsanyi:2012uq}, Fig.~\ref{fig_pbp_check} (left).
Another study by the same group
compared the condensate with overlap fermions (that provide exact
chiral symmetry at non-zero lattice spacing) again to staggered calculations at
$m_\pi=350$~MeV~\cite{Borsanyi:2012xf}. Although the continuum limit for overlap fermions
has not been taken, it seems, Fig.~\ref{fig_pbp_check} (middle),
that disagreement would be unlikely.
The HotQCD collaboration pursued simulations with
domain-wall fermions at $m_\pi=200$~MeV~\cite{:2012jaa}.
A comparison of the staggered
and domain-wall disconnected chiral susceptibility is shown in 
Fig.~\ref{fig_pbp_check} (right). Note
similar location of the peak in the domain-wall (DWF) and HISQ/tree
$N_\tau=12$ data. (Strictly speaking, results for different
lattice actions should be compared in the continuum limit, however,
Fig.~\ref{fig_pbp_check} (right) provides a very encouraging check.)

\section{Deconfinement and fluctuations of conserved charges}

\begin{figure}[h]
\begin{tabular}{ccc}
\parbox{0.31\textwidth}{
  \includegraphics[width=0.32\textwidth]{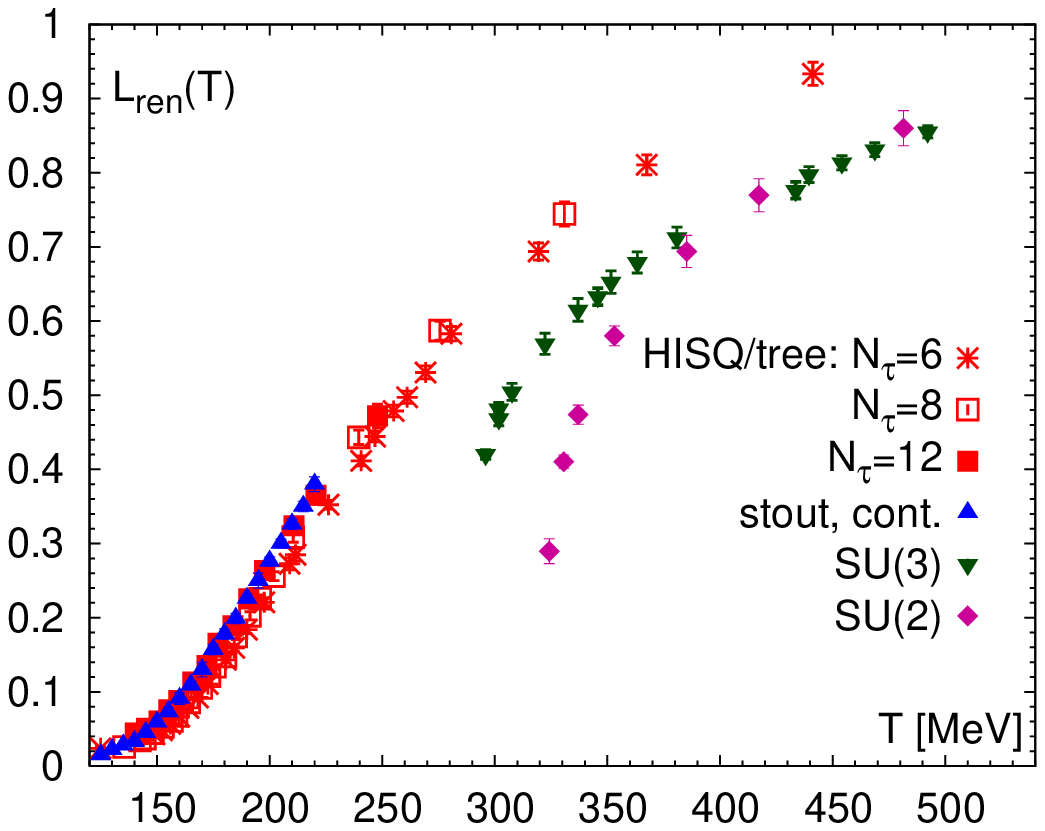}
}
&
\parbox{0.31\textwidth}{
  \includegraphics[width=0.32\textwidth]{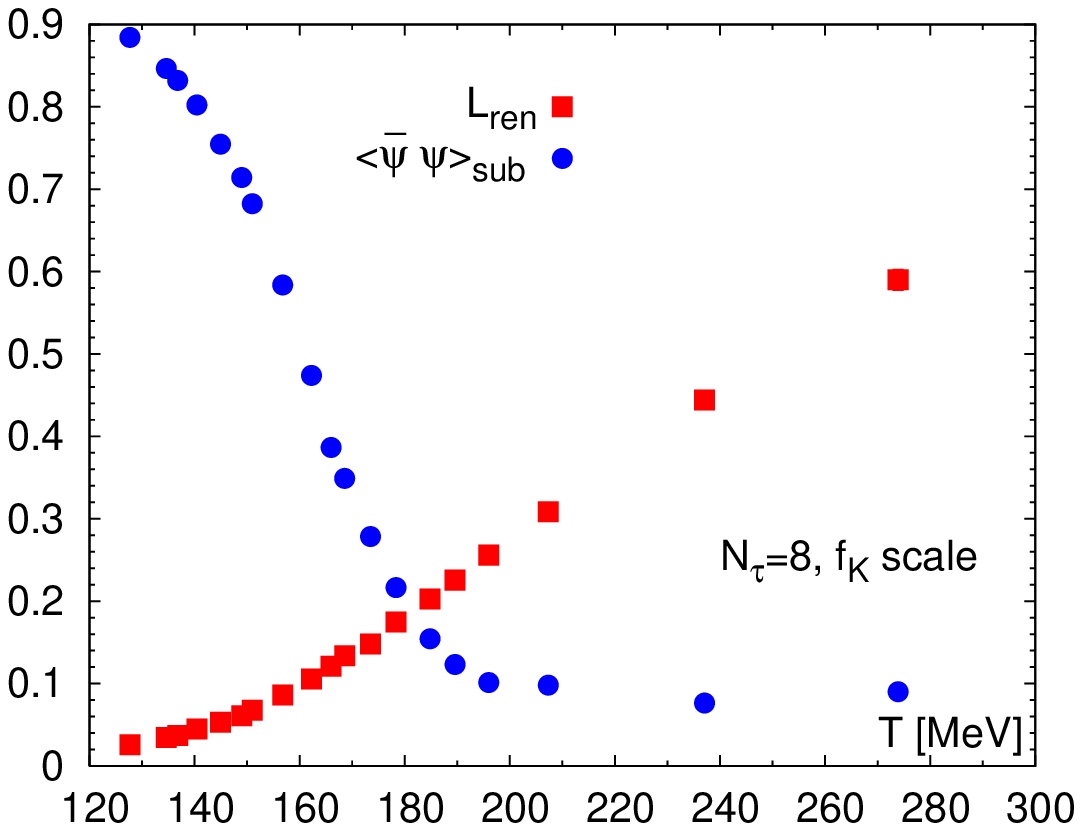}
}
&
\parbox{0.31\textwidth}{
  \includegraphics[width=0.32\textwidth]{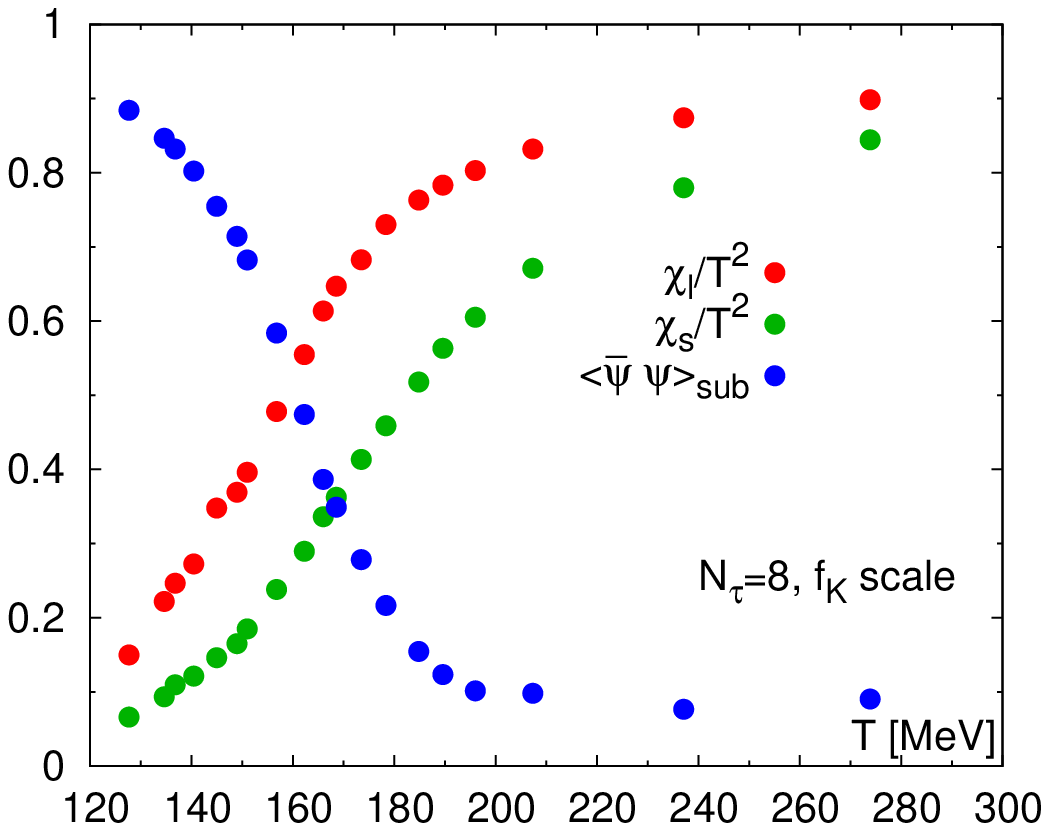}
}
\end{tabular}
\caption{\label{fig_deconf}Temperature dependence of the
Polyakov loop in $SU(2)$ and $SU(3)$ pure gauge theory and QCD (left).
Comparison of the behavior of the renormalized light quark condensate and
the Polyakov loop (middle) and light and strange quark number fluctuations
(right).}
\vspace{-0.6cm}
\end{figure}

QCD with infinitely heavy quarks, {\it i.e.}, $SU(3)$ pure gauge theory, 
exhibits a first-order phase transition at $T_c\simeq260$~MeV~\cite{Boyd:1996bx}.
The order parameter is the renormalized Polyakov loop:
\begin{equation}
L_{ren}(T) = \exp(-c N_{\tau}/2) \left \langle \frac{1}{3} 
{\rm Tr} \prod_{x_0=1}^{N_\tau} U_0(x_0,\vec{x}) \right \rangle,
\label{Ldef}
\end{equation}
where $U_0(x_0,\vec{x})$ is the gauge field in the time direction.
The exponential prefactor takes care of the multiplicative renormalization.
$L_{ren}$ can be related to the free energy $F_Q(T)$ of a static test quark as
$L_{ren}(T)=\exp(-F_Q/T)$.
In pure gauge theory
the Svetitsky-Yaffe argument~\cite{Svetitsky:1982gs} allows to interpret
deconfinement as a transition from the symmetric phase, $L_{ren}=0$, into
the broken phase, $L_{ren}\neq0$, where the Polyakov loop assumes
one of the three equally possible values, related by $Z_3$ symmetry.
So, historically, based on the experience from pure gauge theory,
the Polyakov loop served as an indicator of
deconfinement in lattice QCD. Dialing quark masses to smaller values 
breaks the $Z_3$ symmetry (making one of the values of 
$L_{ren}$ preferable over others, just like
a magnetic field in spin systems). Due to string breaking
by light dynamical quarks in QCD the value of $L_{ren}$ is
always non-zero in the confined phase.
Thus, in QCD the Polyakov
loop loses its meaning of the order parameter and is not related to
any singularity of the partition function (in the chiral limit).
As one can see in Fig.~\ref{fig_deconf} (left)
the temperature dependence of the Polyakov loop in $SU(2)$
and $SU(3)$ pure gauge theory and in QCD is quite different. Sharp behavior in
the former case gives way to smoother change over wide temperature
range. The behavior of the Polyakov loop in QCD at 
low temperatures  can be understood in terms of the hadronic degrees
of freedom ~\cite{Megias:2012kb,Bazavov:2013yv}. However, this description 
breaks down in the vicinity of the transition region. Its role as indicator
for deconfinement thus is obscured.

\begin{figure}[h]
\begin{tabular}{ccc}
\parbox{0.31\textwidth}{
  \includegraphics[width=0.32\textwidth]{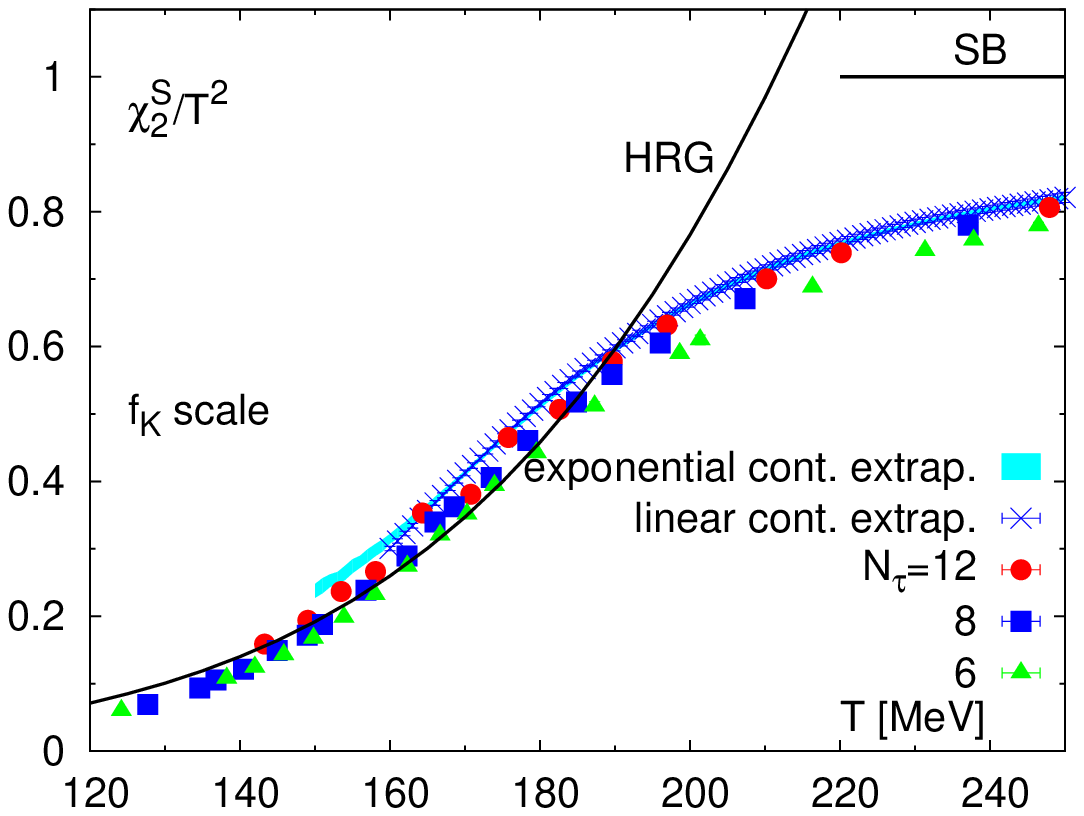}
}
&
\parbox{0.31\textwidth}{
  \includegraphics[width=0.32\textwidth]{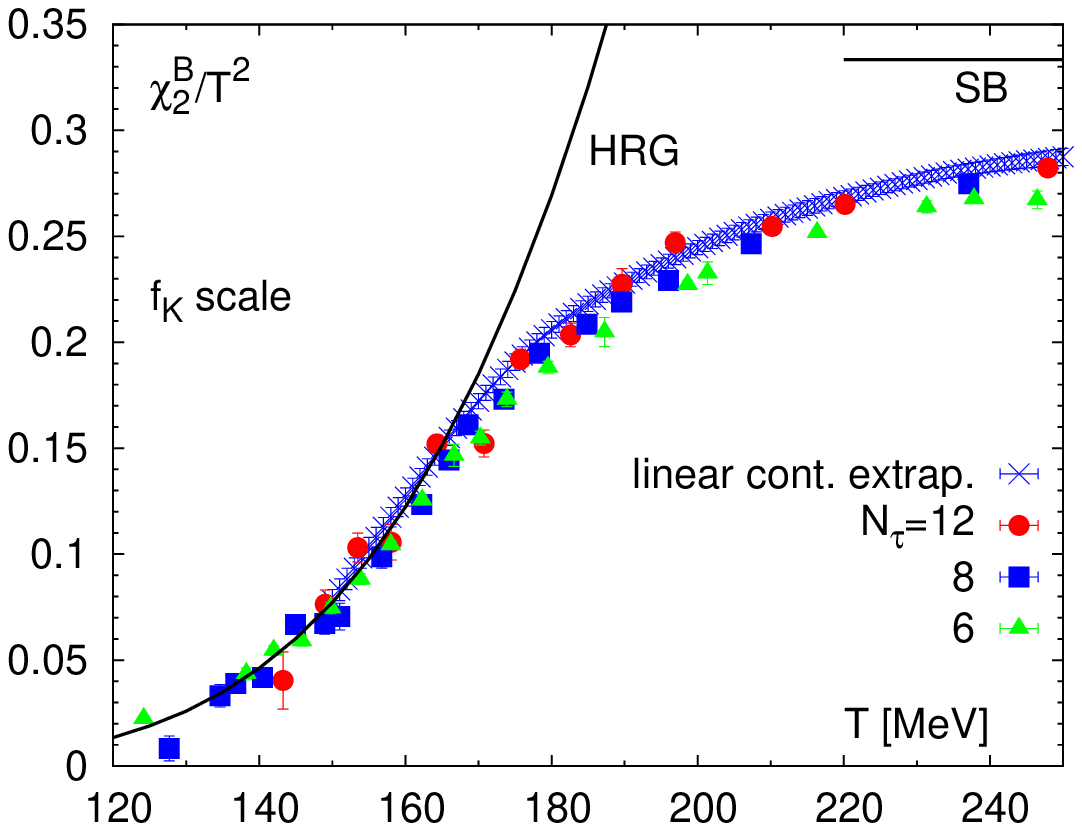}
}
&
\parbox{0.31\textwidth}{
  \includegraphics[width=0.32\textwidth]{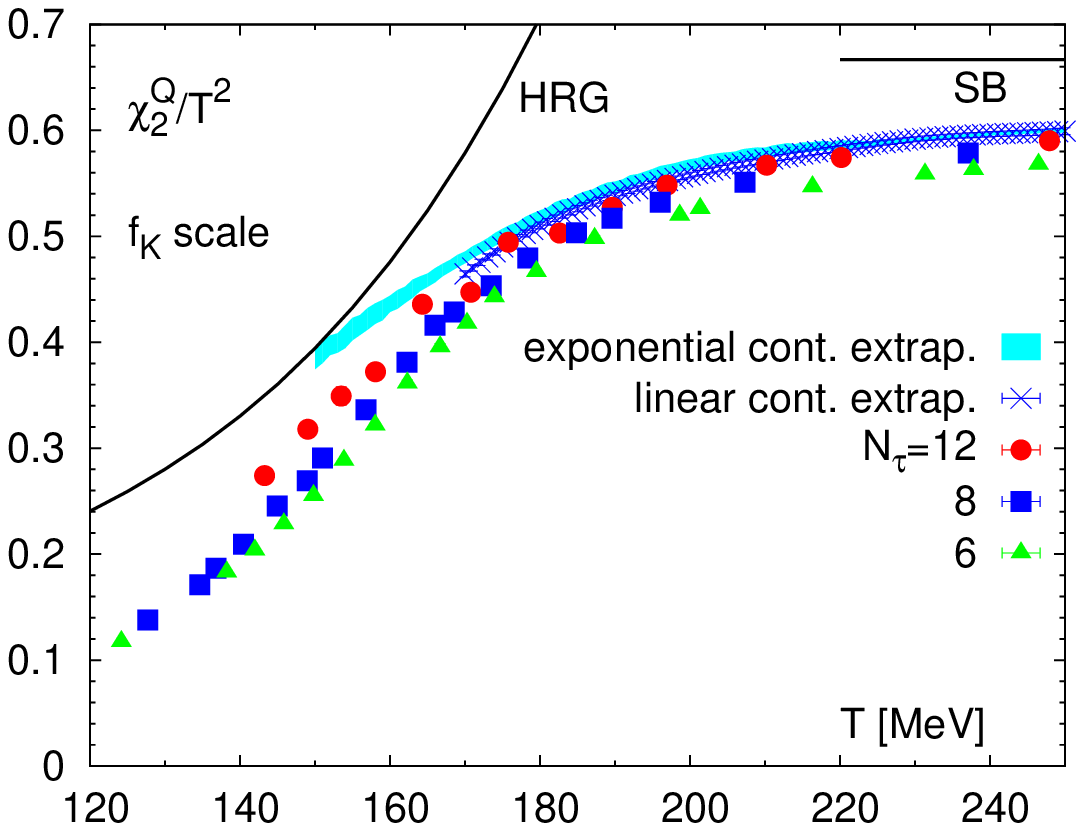}
}
\end{tabular}
\caption{\label{fig_fluc}Fluctuations of strangeness, baryon number and net
electric charge with the HISQ/tree action extrapolated to the continuum and
compared to the Hadron Resonance Gas model.}
\end{figure}

Another possibility to probe deconfinement is through fluctuations
of conserved charges. For instance, light and strange quark number
fluctuations, that are suppressed at low temperature, signal
liberation of the degrees of freedom with quantum numbers of quarks.
In Fig.~\ref{fig_deconf} (middle, right)
we compare the light quark condensate to the
Polyakov loop, light and strange quark number susceptibility.
As one can see, the observables associated with deconfinement rise
somewhat slowly and over substantial temperature range.
This reflects that singular terms contributing to quadratic
fluctuations are still subleading and only the dominant temperature
dependence of regular terms is seen. The analysis of the temperature
dependence of higher order, \textit{e.g.} fourth order, susceptibilities is
needed to become sensitive to deconfining features.
Thus in full QCD associating
a particular temperature with deconfinement by searching for inflection points
or alike in the Polyakov loop or quadratic fluctuations does not appear meaningful.
Rather, full temperature dependence
of various observables should be studied.
In particular, one can study fluctuations and correlations
of baryon number $B$, electric charge $Q$ and strangeness $S$ that
are given by the derivatives of the pressure with respect to
the corresponding chemical potentials
\begin{equation}
\chi_{ijk}^{BQS}=
\frac{\partial^{\,i+j+k}(p/T^4)}{\partial(\mu_{B}/T)^{i}\partial(\mu_{Q}/T)^{j}\partial(\mu_{S}/T)^{k}}.
\label{chi_BQS}
\end{equation}
The fluctuations of strangeness, baryon number and electric
charge, calculated with the HISQ/tree action and extrapolated
to the continuum, are shown in Fig.~\ref{fig_fluc}
(HotQCD collaboration~\cite{Bazavov:2012jq}).
Solid curves represent
the Hadron Resonance Gas model, which is in remarkable agreement
with the lattice data up to $T\sim 150-160$~MeV. (Electric charge
fluctuations are the most sensitive to the cutoff effects in the pion
sector and obtaining the continuum limit is, thus, a more subtle issue.)
The results from the Budapest-Wuppertal collaboration~\cite{Borsanyi:2011sw}
for the same quantities are in complete agreement.

\begin{figure}[h]
\begin{minipage}{0.32\textwidth}
\includegraphics[width=\textwidth]{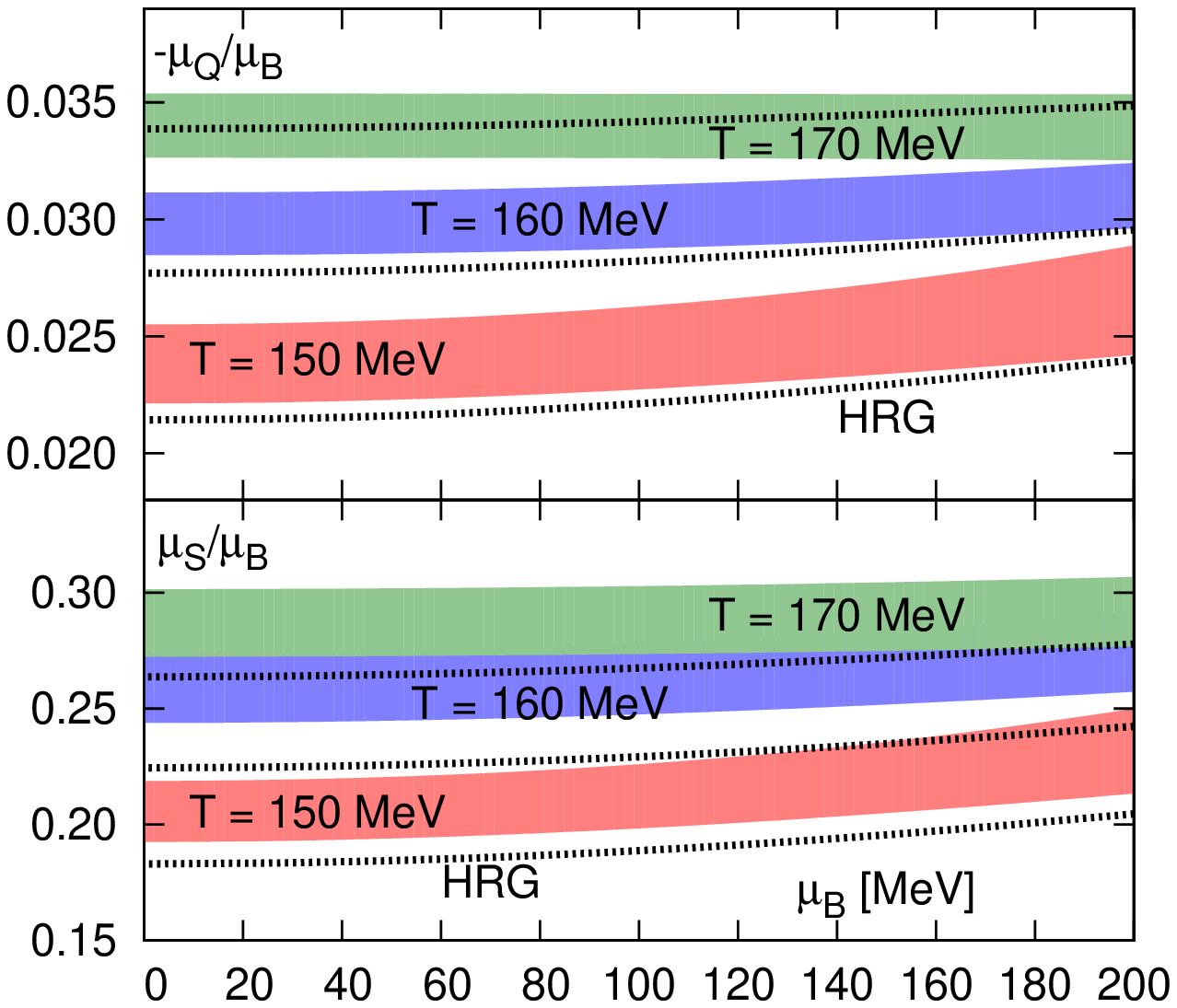}
\vspace{-0.7cm}
\caption{\label{fig_muQ_muS}Strangeness and electric charge chemical potential
at several temperatures with the constraints (\ref{mu_constr}).}
\end{minipage}\hfill
\begin{minipage}{0.32\textwidth}
\includegraphics[width=\textwidth]{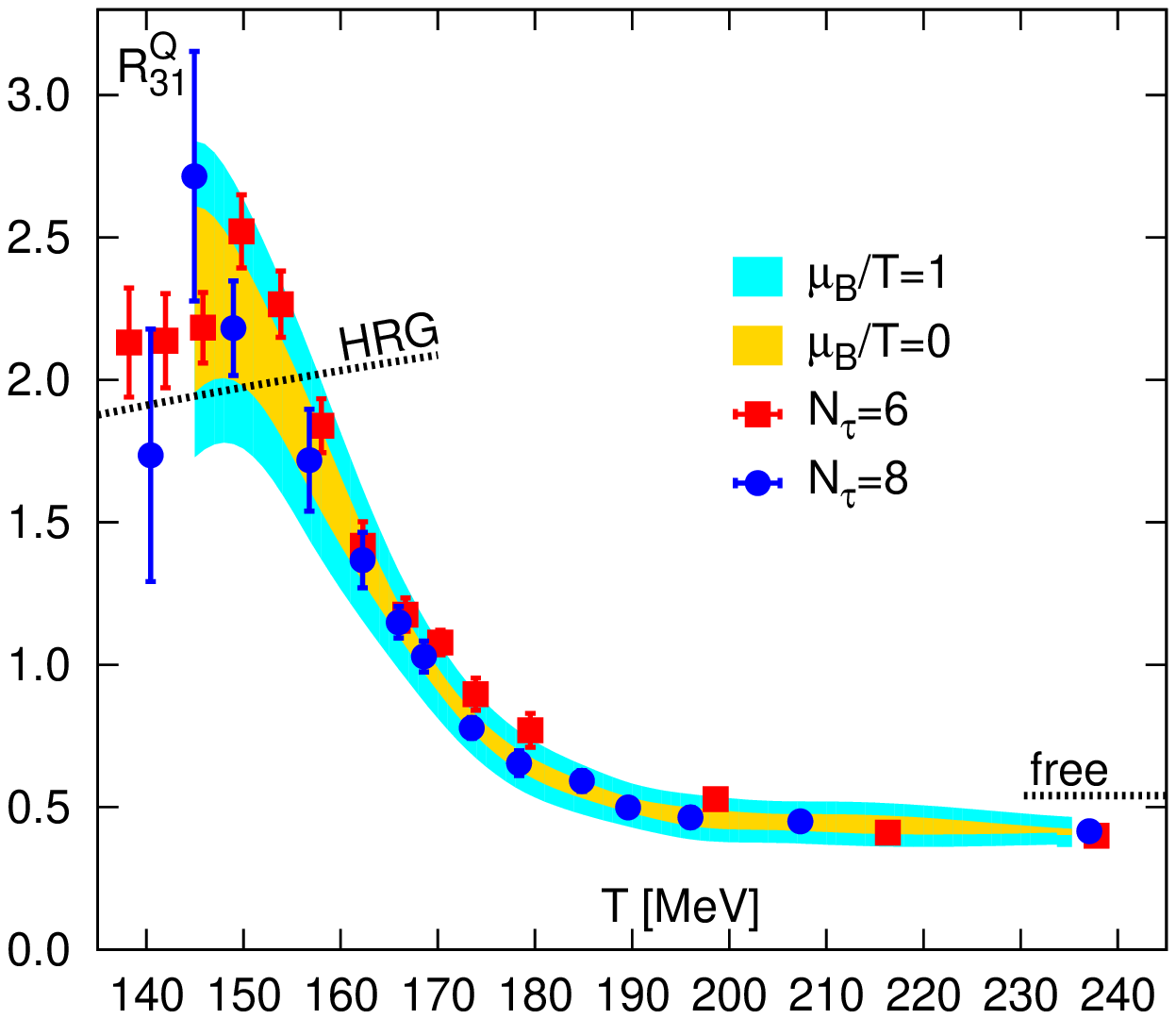}
\vspace{-0.7cm}
\caption{\label{fig_R31}The dependence $R_{31}^Q(T)$ allows
for determination of the freeze-out temperature $T_f$.}
\end{minipage}\hfill
\begin{minipage}{0.32\textwidth}
\includegraphics[width=\textwidth]{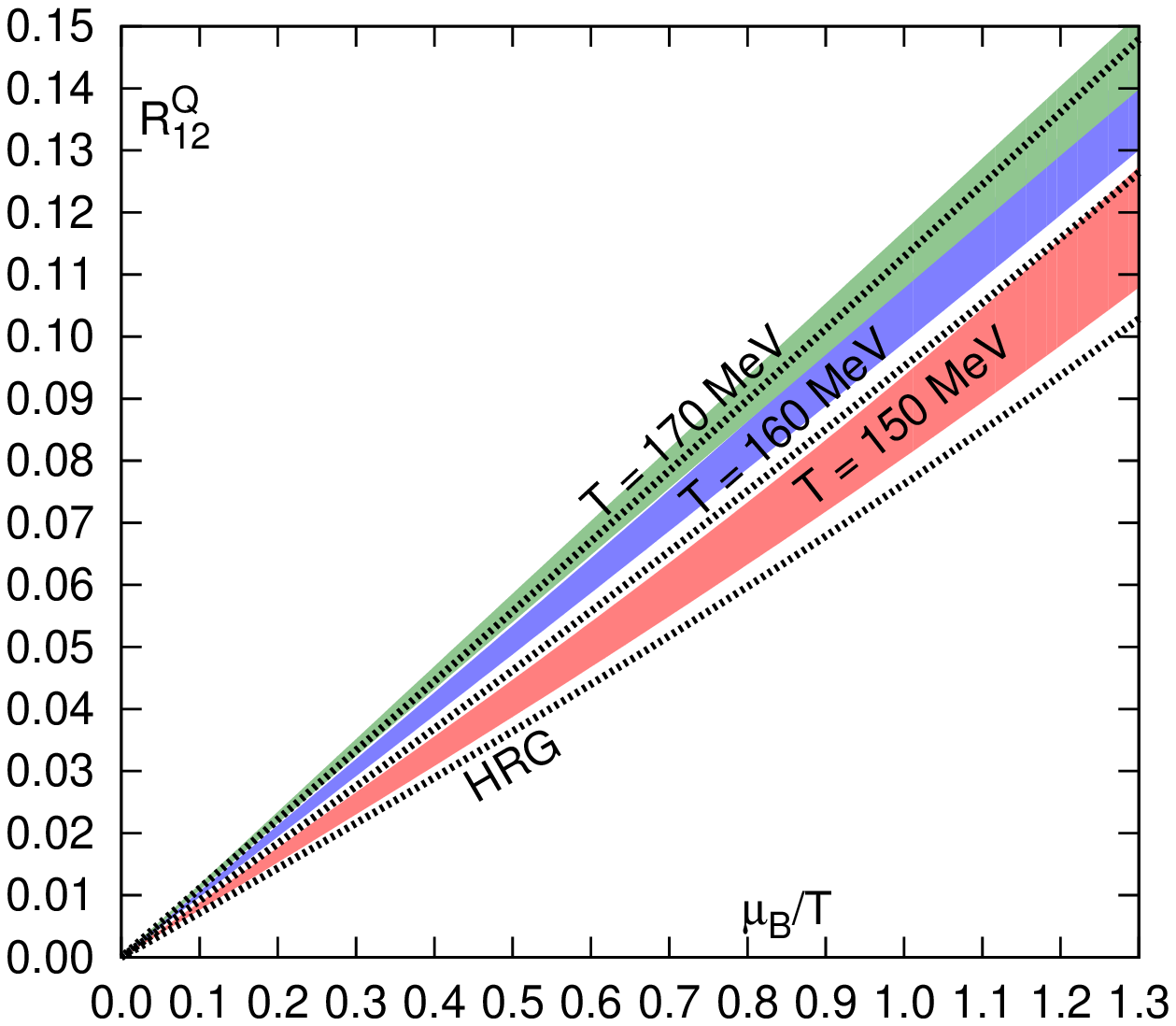}
\vspace{-0.7cm}
\caption{\label{fig_R12}The dependence $R_{12}^Q(T)$ at given $T_f$ determines
the freeze-out chemical potential $\mu_B^f$.}
\end{minipage} 
\end{figure}

To probe the critical behavior of QCD further one can consider higher-order
cumulants of conserved charges. Comparing ratios of cumulants calculated on
the lattice with experimental data, under assumption that at the time
of chemical freeze-out the system created in heavy-ion collisions can
be described by equilibrium thermodynamics with temperature $T_f$ and
baryon chemical potential $\mu_B^f$, one can extract these freeze-out
parameters. A recent attempt to use electric charge fluctuations for
this purpose has been presented in Ref.~\cite{Bazavov:2012vg}.
By constraining the mean net strangeness and electric charge
to those of the incident nuclei,
\begin{equation}
M_S=0,\,\,\,M_Q=rM_B,\,\,\,(r\simeq0.4\mbox{ in gold-gold and lead-lead collisions}),
\label{mu_constr}
\end{equation}
one can determine the strange and electric charge chemical potential of the system.
To the next-to-leading order (NLO) in the baryon chemical potential:
\begin{equation}
\frac{\mu_Q}{T} = q_1\ \frac{\mu_B}{T} + q_3\ \left(\frac{\mu_B}{T}\right)^3,\,\,\,\,\,
\frac{\mu_S}{T} = s_1\ \frac{\mu_B}{T} + s_3\ \left(\frac{\mu_B}{T}\right)^3.
\label{mu_expansion}
\end{equation}
The $q_i, s_i$ coefficients are related to ratios of generalized susceptibilities,
\textit{i.e.} various combinations of derivatives of the partition function with
respect to chemical potentials, Eq.~(\ref{chi_BQS}).
It has been shown in Ref.~\cite{Bazavov:2012vg} that
NLO corrections are small and at NLO the dependence of $\mu_Q/\mu_B$ and 
$\mu_S/\mu_B$ on $\mu_B$ at several temperatures is shown in Fig.~\ref{fig_muQ_muS}.
The bands
give total uncertainty of the lattice calculation and dashed lines represent the
HRG model. Once $\mu_Q$ and $\mu_S$ satisfying the constraint (\ref{mu_constr}) are
fixed, ratios of cumulants can be evaluated, and Ref.~\cite{Bazavov:2012vg} focused on
\begin{eqnarray}
R_{12}^X &\equiv& 
\frac{M_X}{\sigma_X^2} = 
\frac{\mu_B}{T} \left(
R_{12}^{X,1} + R_{12}^{X,3}\ \left(\frac{\mu_B}{T}\right)^2 + {\cal O}((\mu_B/T)^4)
\right)\; ,
\label{R12} \\
R_{31}^X &\equiv& 
\frac{S_X \sigma_X^3}{M_X} = 
 R_{31}^{X,0} + R_{31}^{X,2}\ \left(\frac{\mu_B}{T}\right)^2 + {\cal O}((\mu_B/T)^4)\; ,
\label{R31}
\end{eqnarray}
where $X=B$ or $Q$, $M_X$ is mean, $\sigma_X^2$ is variance and $S_X$ skewness for corresponding
conserved charges. The results are shown in Fig.~\ref{fig_R31} and \ref{fig_R12}.
If $R_{31}^Q$ is determined from
experiment, one can use the band in Fig.~\ref{fig_R31} to convert it 
to the freeze-out temperature $T_f$.
Then from an experimentally determined ratio $R_{12}^Q$, $T_f$ and the corresponding band in
Fig.~\ref{fig_R12} one can find the freeze-out chemical potential $\mu_B^f$.
Once the full experimental analysis of $R_{31}^Q$ and $R_{12}^Q$ is
available, the results of Ref.~\cite{Bazavov:2012vg} can determine $T_f$ and $\mu_B^f$ for
various center-of-mass energies and thus map a part of the freeze-out curve on the
QCD phase diagram.

\section{A list of other results on the lattice}

Due to the space constraints many other recent interesting
calculations on the lattice could not be included. Among those
I would like to mention:
\begin{itemize}
\item Progress on the equation of state with 2+1 flavors of
staggered fermions: an update of the previous results from the
Budapest-Wuppertal~\cite{Borsanyi:2012vn} and
HotQCD \cite{Petreczky:2012gi,Bazavov:2012bp} Collaboration.
Preliminary results still indicate a discrepancy in $150-300$~MeV range,
which may be resolved by ongoing calculations.
\item 2+1 flavor equation of state with Wilson fermions
from the WHOT-QCD collaboration~\cite{Umeda:2012nn}.
\item The equation of state with dynamical charm quark, see
Refs.~\cite{Borsanyi:2012vn} and \cite{Bazavov:2012kf}
for preliminary results.
\item Calculation of the charmonium screening masses
from spatial correlation functions in 2+1 flavor QCD~\cite{Karsch:2012na} and
Maximum Entropy Method (MEM) reconstruction of the charmonium spectral functions in
quenched QCD~\cite{Ding:2012sp}.
\item Thermodynamics of the crossover region with high
magnetic fields~\cite{Bali:2012zg}.
\end{itemize}

\section*{Acknowledgments}
I would like to thank the organizers of Hot Quarks 2012 for a very
productive and interesting meeting and Frithjof Karsch and 
Peter Petreczky for
careful reading and comments on the manuscript. This work was in
part supported by U.S. Department of Energy under
Contract No. DE-AC02-98CH10886.


\section*{References}
\begin{thebibliography}{99}

\bibitem{Bernard:2004je} 
  Bernard C {\it et al.}  [MILC Collaboration] 2005
  {\it Phys. Rev.} D {\bf 71} 034504
  ({\it Preprint} hep-lat/0405029)

\bibitem{Cheng:2006qk} 
  Cheng M, Christ N H, Datta S, Van der Heide J 
  {\it et al.}
  2006 {\it Phys. Rev.} D {\bf 74} 054507
  ({\it Preprint} hep-lat/0608013)

\bibitem{Aoki:2006we} 
  Aoki Y, Endrodi G, Fodor Z, Katz S D and Szabo K K 2006
  {\it Nature} {\bf 443} 675
  ({\it Preprint} hep-lat/0611014)

\bibitem{Petreczky:2012rq}
  Petreczky P 2012
  {\it J. Phys.} G {\bf 39} 093002
  ({\it Preprint} 1203.5320)

\bibitem{Pisarski:1983ms} 
  Pisarski R D and Wilczek F 1984
  {\it Phys. Rev.} D {\bf 29} 338

\bibitem{Ejiri:2009ac} 
  Ejiri S, Karsch F, Laermann E, Miao C {\it et al.} 2009
  {\it Phys. Rev.} D {\bf 80} 094505
  ({\it Preprint} 0909.5122)

\bibitem{Bazavov:2011nk} 
  Bazavov A, Bhattacharya T, Cheng M, DeTar C {\it et al.} [HotQCD Collaboration] 2012
  {\it Phys. Rev.} D {\bf 85} 054503
  ({\it Preprint} 1111.1710)

\bibitem{Borsanyi:2010bp} 
  Borsanyi S {\it et al.}  [Wuppertal-Budapest Collaboration] 2010
  {\it JHEP} {\bf 1009} 073
  ({\it Preprint} 1005.3508)

\bibitem{Bazavov:2010pg} 
  Bazavov A and Petreczky P 2010
  {\it PoS} {\bf LATTICE2010} 169
  ({\it Preprint} 1012.1257)

\bibitem{Borsanyi:2012uq} 
  Borsanyi S, Durr S, Fodor Z, Hoelbling C {\it et al.} 2012
  {\it JHEP} {\bf 1208} 126
  ({\it Preprint} 1205.0440)

\bibitem{Borsanyi:2012xf} 
  Borsanyi S, Delgado Y, Durr S, Fodor Z {\it et al.} 2012
  {\it Phys. Lett.} B {\bf 713} 342
  ({\it Preprint} 1204.4089)

\bibitem{:2012jaa} 
  Bazavov A, Bhattacharya T, Buchoff M I, Cheng M {\it et al.} [HotQCD Collaboration] 2012
  {\it Phys. Rev.} D {\bf 86} 094503
  ({\it Preprint} 1205.3535)

\bibitem{Boyd:1996bx} 
  Boyd G, Engels J, Karsch F, Laermann E {\it et al} 1996
  {\it Nucl. Phys.} B {\bf 469} 419
  ({\it Preprint} hep-lat/9602007)

\bibitem{Svetitsky:1982gs} 
  Svetitsky B and Yaffe L G 1982
  {\it Nucl. Phys.} B {\bf 210} 423

\bibitem{Megias:2012kb} 
  Megias E, Ruiz Arriola E and Salcedo L L 2012
  {\it Phys. Rev. Lett.}  {\bf 109} 151601
  ({\it Preprint} 1204.2424)

\bibitem{Bazavov:2013yv} 
  Bazavov A and Petreczky P 2013
  ({\it Preprint} 1301.3943)

\bibitem{Bazavov:2012jq} 
  Bazavov A, Bhattacharya T, DeTar C E, Ding H-T {\it et al.}  [HotQCD Collaboration] 2012
  {\it Phys. Rev.} D {\bf 86} 034509
  ({\it Preprint} 1203.0784)

\bibitem{Borsanyi:2011sw} 
  Borsanyi S, Fodor Z, Katz S D, Krieg S, Ratti C and Szabo K 2012
  {\it JHEP} {\bf 1201}, 138
  ({\it Preprint} 1112.4416)

\bibitem{Bazavov:2012vg} 
  Bazavov A, Ding H T, Hegde P, Kaczmarek O {\it et al.} 2012
  {\it Phys. Rev. Lett.}  {\bf 109} 192302
  ({\it Preprint} 1208.1220)

\bibitem{Borsanyi:2012vn} 
  Borsanyi S, Endrodi G, Fodor Z, Katz S D {\it et al.} 2011
  {\it PoS} {\bf LATTICE2011} 201
  ({\it Preprint} 1204.0995)

\bibitem{Petreczky:2012gi} 
  Petreczky P [HotQCD Collaboration] 2012
  {\it PoS} {\bf LATTICE2012} 069
  ({\it Preprint} 1211.1678)

\bibitem{Bazavov:2012bp} 
  Bazavov A [HotQCD Collaboration] 2012
  ({\it Preprint} 1210.6312)

\bibitem{Umeda:2012nn} 
  Umeda T {\it et al.} [WHOT-QCD Collaboration] 2012
  {\it PoS} {\bf LATTICE2012} 074
  ({\it Preprint} 1212.1215)

\bibitem{Bazavov:2012kf}
  Bazavov A {\it et al.} [MILC Collaboration] 2012
  {\it PoS} {\bf LATTICE2012} 071

\bibitem{Karsch:2012na} 
  Karsch F, Laermann E, Mukherjee S and Petreczky P 2012
  {\it Phys. Rev.} D {\bf 85} 114501
  ({\it Preprint} 1203.3770)

\bibitem{Ding:2012sp} 
  Ding H T, Francis A, Kaczmarek O, Karsch F, Satz H and Soeldner W 2012
  {\it Phys. Rev.} D {\bf 86} 014509
  ({\it Preprint} 1204.4945)

\bibitem{Bali:2012zg} 
  Bali G S, Bruckmann F, Endrodi G, Fodor Z, Katz S D and Schafer A
  {\it Phys. Rev.} D {\bf 86} 071502
  ({\it Preprint} 1206.4205)

\end{thebibliography}
\end{document}